\documentclass[journal]{IEEEtran}
\ifCLASSINFOpdf \else \fi
\usepackage{url}
\usepackage{color}
\usepackage{color,array}
\usepackage{lipsum}
\usepackage{hhline}
\usepackage{yfonts}
\usepackage{stackengine}
\usepackage[table,xcdraw]{xcolor}
\usepackage{amssymb,amsmath,color,graphicx, amsbsy}
\usepackage{graphicx}
\newcolumntype{P}[1]{>{\centering\arraybackslash}p{#1}}
\newcolumntype{M}[1]{>{\centering\arraybackslash}m{#1}}
\usepackage{cite}
\usepackage{algorithm}
\usepackage{algorithmic}
\usepackage{amsmath}
\usepackage{amsthm}
\usepackage{amsfonts}
\usepackage{amssymb}
\usepackage{graphicx}
\usepackage{epstopdf}
\usepackage{makeidx}
\usepackage{outlines}
\usepackage[noprefix]{nomencl}
\usepackage{textcomp}
\usepackage{soul}
\usepackage{multirow}
\usepackage{bm}
\newcolumntype{P}[1]{>{\centering\arraybackslash}p{#1}}
\newcolumntype{M}[1]{>{\centering\arraybackslash}m{#1}}
\usepackage{xcolor}
\usepackage[table]{xcolor}

\usepackage{array}
\usepackage{tabularx}
\usepackage{hyperref}
\hypersetup{
	colorlinks=true,
	linkcolor=black,
	filecolor=black,
	urlcolor=black,
	anchorcolor	=black,
	citecolor=black,
}

\usepackage{graphicx}
\usepackage{amssymb}
\usepackage{amsmath}
\usepackage{longtable}
\usepackage{algorithmic}
\bstctlcite{IEEEexample:BSTcontrol}
\usepackage{array}

\ifCLASSOPTIONcompsoc
 \usepackage[caption=false,font=normalsize,labelfont=sf,textfont=sf]{subfig}
\else
 \usepackage[caption=false,font=footnotesize]{subfig}
\fi
\usepackage{url}
\usepackage{enumitem}
\usepackage{lettrine}
\usepackage{cite}
\usepackage{color,soul,graphics,graphicx,amsmath,amsfonts,amssymb,times,setspace,cite,fancyhdr,epstopdf}
\usepackage{booktabs}
\usepackage{multirow}
\usepackage{tabularx}
\usepackage[table,xcdraw]{xcolor}
\usepackage{adjustbox}
\usepackage{float}
\usepackage{longtable}
\allowdisplaybreaks

\DeclareMathSymbol{\N}{\mathbin}{AMSb}{"4E}
\DeclareMathSymbol{\Z}{\mathbin}{AMSb}{"5A}
\DeclareMathSymbol{\R}{\mathbin}{AMSb}{"52}
\DeclareMathSymbol{\Q}{\mathbin}{AMSb}{"51}
\DeclareMathSymbol{\I}{\mathbin}{AMSb}{"49}
\DeclareMathSymbol{\C}{\mathbin}{AMSb}{"43}
\DeclareMathSymbol{\Exp}{\mathbin}{AMSb}{"45}
\DeclareMathSymbol{\Prob}{\mathbin}{AMSb}{"50}
\usepackage{comment}

\ifCLASSINFOpdf
  
\else
 
\fi

\ifCLASSOPTIONcompsoc
 \usepackage[caption=false,font=normalsize,labelfont=sf,textfont=sf]{subfig}
\else
 \usepackage[caption=false,font=footnotesize]{subfig}
\fi

\renewenvironment{longtable}{\begin{center}\begin{tabular}}{\end{tabular}\end{center}}

\renewcommand{\toprule}[2]{\hline}
\renewcommand{\midrule}[2]{\hline}
\renewcommand{\bottomrule}[2]{\hline}

\begin{document}

\title{The Dilemma of Electricity Grid Expansion Planning in Areas at the Risk of Wildfire}

\author{Muhammad Waseem,~\IEEEmembership{Student Member,~IEEE},
        Reza  Bayani,~\IEEEmembership{Student Member,~IEEE}, and
        Saeed D. Manshadi,~\IEEEmembership{Member,~IEEE}
        \thanks{The authors are with the Department of Electrical and Computer Engineering, San Diego State University, San Dieg, CA, 92182, USA email:(mhuwaseem@gmail.com; rbayani@sdsu.edu; smanshadi@sdsu.edu)}
}

\markboth{}%
{Shell \MakeLowercase{\textit{et al.}}: Bare Demo of IEEEtran.cls for IEEE Journals}

\maketitle

{
\begin{abstract} 
The utilities consider \emph{public safety power shut-offs} imperative for the mitigation of wildfire risk. This paper presents expansion planning of power system under fire hazard weather conditions.
The power lines are quantified based on the risk of fire ignition. A 10-year expansion planning scenario is discussed to supply power to customers 
by considering three decision variables: distributed solar generation; modification of existing power lines; addition of new lines. Two-stage robust optimization problem is formulated and solved using Column-and-Constraint Generation Algorithm to find improved balance among de-energization of customers, distributed solar generation, modification of power lines, and addition of new lines. It involves lines de-energization of high wildfire risk regions and serving the customers by integrating distributed solar generation. The impact of de-energization of lines on distributed solar generation is assessed. The number of hours each line is energized and total load shedding during a 10-year period is evaluated. Different uncertainty levels for system demand and solar energy integration are considered to find the impact on the total operation cost of the system. The effectiveness of the presented algorithm is evaluated on 6- and 118-bus systems.
\bstctlcite{IEEEexample:BSTcontrol}
\end{abstract}
}
\begin{IEEEkeywords}
Wildfire, Electricity grid expansion planning, Wind speed, Wind gust, Two-stage robust optimization, Column-and-constraint generation algorithm
\end{IEEEkeywords}

\IEEEpeerreviewmaketitle
\vspace{-.25cm}
\subsection*{Indices}
\vspace{-0.3cm}
\begin{longtable}{l p{2.7in}} 
\textit{g} $\in$ \(\mathcal{G}\) & Index of generator in the set of generators\\
\textit{i} $\in$ \(\mathcal{I}\) & Index of bus in the set of buses\\
\textit{l} $\in$ \(\mathcal{L}\) & Index of line in the set of power lines\\
$\mathcal{L}^{fr}_{i}(\mathcal{L}^{to}_{i})$ & Set of lines originated from (destined to) bus $i$\\
\textit{s} $\in$ \(\mathcal{S}\) & Index of scenario in the set of scenarios\\
\textit{y} $\in$ \(\mathcal{Y}\) & Index of year in the set of years\\
$z \in \mathcal{Z}$ & Index for segment of generating unit cost curve
\end{longtable}
\vspace{-0.2cm}
\subsection*{Binary Decision Variables}
\vspace{-0.2cm}
\begin{longtable}{l p{2.6in}} 
\textit{\(I_{s,y}^l\)} & Energization/de-energization status of line $l$ during scenario $s$ and year $y$\\
\textit{\(I_{l,y}^{exs}\)} & Indicator for physical existence of line $l$ at year $y$\\
\textit{\(I_{l,y}^{mod}\)} & Indicator for modification of line $l$ at year $y$\\
\textit{\(u_{s,y}^{D,i}\), \(v_{s,y}^{D,i}\)} & Uncertainties of bus $i$ during year $y$ \\
\textit{\(u_{s,y}^{R,i}\), \(v_{s,y}^{R,i}\)} & Solar generation uncertainties during year $y$
\end{longtable}

\subsection*{Continuous Decision Variables}
\begin{longtable}{l p{2.8in}} 
\textit{\(aux_{s,y}^l\)} & Auxiliary variable denoting decision for modification of existing lines during scenario $s$ and year $y$\\
\textit{\(P_{s,y}^{D,i}\)} & Scheduled consumer demand at bus $i$ during scenario $s$ and year $y$\\
\textit{\(P_{s,y}^{d,i}\)} & Demand served at bus $i$ during scenario $s$ and year $y$\\
\textit{\(P_{s,y}^{L,l}\)} & Real power flowing at line $l$ during scenario $s$ and year $y$\\
\textit{\(P_{s,y}^{G,g}\)} & Dispatched power of generator $g$ during scenario $s$ and year $y$\\
\textit{\(P_{s,y}^{R,i}\)} & Dispatch of solar generation unit at bus $i$ during scenario $s$ and year $y$\\
\textit{\(\overline{P}_{y}^{R,i}\)} & Installed solar generation capacity at bus $i$ during year $y$\\
\textit{\(P_{s,y}^{g,z}\)} & Share of segment $z$ in power generated by unit $g$ during scenario $s$ and year $y$\\
\textit{\(SC_{s,y}^{l}\)} & Fire ignition score of energized line $l$ during scenario $s$ and year $y$\\
\textit{\(\theta_{s,y}^{i}\)} & Voltage angle at bus $i$ during scenario $s$ and year $y$\\
\textit{\(\psi_{s,y}^{l}\)} & Fire ignition score of line $l$ during scenario $s$ and year $y$\\
\textit{\(\kappa_{s,y}^{i}\)} & Availability of solar generation at bus $i$ during scenario $s$ and year $y$ 
\end{longtable}
\subsection*{Parameters}
\vspace{-0.3cm}
\begin{longtable}{l p{2.85in}} 
\textit{\(c_g^{z}\)} & Generation cost of segment $z$ and unit $g$ \\
\textit{\({CN_y^l}\)} & Cost of new line $l$ installment during year $y$ \\
\textit{\({CD_y}\)} & Cost of DER installment during year $y$ \\
\textit{\({CM_y^l}\)} & Cost of modification of line $l$ during year $y$\\
\textit{E} & Budget of uncertainty\\
\textit{\( K, M\)} & $K$ is a penalty factor for not serving load and $M$ is an arbitrary large constant\\
\textit{\({R_s}\)} & Number of hours in which a scenario $s$ occurs per year \\
\textit{\({x_l}\)} & Reactance of power line $l$\\
\textit{\(\mathcal{\triangle}P_{s,y}^{d,i}\)} & Deviation in demand at bus $i$ during scenario $s$ and year $y$ \\
\textit{\({\mathcal{E}}\)} & Level of risk tolerance\\
\textit{\({\rho^l}\)} & Flag representing the existence of line $l$ : 1 if line exists otherwise 0\\
\textit{\(\delta\)} & A parameter to reduce the fire ignition score when a line is modified
\end{longtable}

\section{Introduction}
\subsection{Motivation}
\lettrine{T}{he} frequency of wildfire events in the past decades has been increasing \cite{weber2020spatiotemporal}. Climate change and higher average temperatures are counted among contributing factors to this phenomenon \cite{jolly2015climate}. Reportedly, power lines are listed as a cause of wildfire ignitions \cite{psps_1}. From 2015-2017, the electric power lines were responsible for 414 fire ignition events in California, as declared by Pacific Gas \& Electric Company (PG\&E) \cite{pge_report}.  More than 4,000 fires occurred during the 4-year period between 2010 and 2014 in Texas which were caused by power lines\cite{texas_lines}. The electric power equipment ignited fires that led to the death of 179 people during 2009 Black Saturday wildfire in Australia \cite{teague2010final}. The fires initiated by power lines are known to be more extensive \cite{miller2017electrically_1, keeley2018historical_1}. A power line triggered California Camp Fire in 2018 which resulted in \$9.3 billion residential property damage and death of 84 people\cite{camp_fire_ref}. Wildfire ignitions by power lines are strongly correlated with the wind speed, as with even mild increases in the wind speed, the probability of failures within power lines is multiplied \cite{mitchell2013power}. The breaking of towers and failure of power lines are among the faults that could occur due to the high wind speeds \cite{dumas2019extreme}. The contact between surrounding vegetation and the conductors is also a frequent cause of fire ignition in power systems \cite{jazebi2019review, benner2019dfa, russell2012distribution}. 
The seasonal Santa Ana winds are associated with ignition and acceleration of several wildfires in California, such as the 2017 Lilac Fire \cite{sdgefireplan}.
In order to prevent fire ignition, utilities in California embrace \emph{public safety power shut-offs} (PSPS) events during extreme wind conditions. Between the period of 2013 to 2020, California utilities have conducted 51 PSPS events \cite{psps_1}.

\par 
A pre-planned operation of power system network is critical for reliable power delivery to the customers. The primary purpose of expansion planning is to ensure power network can keep up with the system-wide changes that can potentially impact the network operations in the future, such as electricity demand growth over the time, changes in the power generation fleet, or even changes in the climate conditions.
Expansion planning has acquired an increased attention due to the growth in renewable energy resources' penetration in power system and de-commissioning of traditional power generation resources \cite{aniti2014investment}. Additionally, the resilient operation and expansion planning of electricity grid are critical to address the issues North American electricity grids are facing as a result of ever-increasing wildfires. 
In this work, an expansion planning problem is proposed to prepare against natural wildfires, by taking the uncertain nature of these phenomenons into consideration.

\par 
Based on our previous work \cite{waseem2021resilient}, we presented a machine learning model to quantify the risk of fire ignition by power lines under fire hazard weather conditions \cite{waseem_1}. A risk-averse resilient operation of electricity grid based on the fire ignition score is presented in \cite{waseem_2}. In continuation to these works, this paper presents a two-stage robust expansion planning problem for a power system by considering wildfire ignition risk. The wildfire ignition risk is a value obtained based on the conductor clashing score determined by considering physical, structural, and meteorological conditions including span of power line, conductor diameter, phase clearance, wind speed, wind gust, and wind direction as determined in our previous work \cite{waseem_1}. The conductor clashing score is calculated based on how many points of the conductors are coming in contact with each other by varying these features and it ranges [0,1), where 0 means no risk of conductor clashing while 0.99 indicates the whole conductor except both ends is at the risk of clashing.
Based on the risk score associated with each power line in the system, the objective is to establish balance between preemptive de-energization of power lines, integration of solar energy resources, and hardening measures. Hardening the power system network against natural disasters refers to physically modifying the infrastructure to make it less vulnerable to severe conditions. Such measures are costly due to their physical nature. In this work, the hardening measures include modification of power lines and addition of new lines. Modification of power line refers to the relocation of a line to an area having lower wildfire ignition risk.
This paper seeks the answers to the following questions. \emph{1) Given a quantified risk of wildfire ignition during different weather conditions, which power lines should be de-energized? 
2) To what extent is the system's operation affected in case of line de-energization? 
3) What are the implications of uncertainties in system demand and solar energy generation on the expansion planning decision variables? 4) How do different uncertainty levels impact the operation cost of the system? 5) How do distributed solar generation, modification of power lines, and addition of new lines help the operator serve customers in a power system at risk of severe wildfire?}

\subsection{Literature Review}

\par Several aspects of the power system expansion planning have been subject of interest. Power system expansion planning considering demand growth has been subject of some research works \cite{khayatian2017integrated,wu2020ac,cesena2015flexible,silva2006transmission,baringo2017stochastic,qiu2014multi}.
A two-stage stochastic mixed-integer optimization problem is used to study the expansion planning in electricity market considering the uncertainty factor of load growth
in \cite{khayatian2017integrated}. A multi-stage AC/DC distribution system expansion planning using mixed-integer linear programming model considering the uncertainty in load demand
is presented in \cite{wu2020ac}. An expansion planning model of multi-energy generation systems under the uncertainty in load demand
is presented in \cite{cesena2015flexible}. The transmission network expansion planning by considering the load demand uncertainty levels is presented in \cite{silva2006transmission}. A generation and transmission expansion planning problem using the stochastic adaptive robust optimization method by considering both short-term and long-term uncertainty levels in peak demand 
is presented in \cite{baringo2017stochastic}.  A multi-stage expansion co-planning of power lines, gas pipelines, and gas power plants is modeled as mixed-integer non-linear programming problem under uncertainty of load growth in \cite{qiu2014multi}.

\par With the rapidly increasing interest in renewable generation, expansion planning of power systems subject to growing penetration of these resources has grabbed research attention \cite{shortt2012accommodating,palmintier2015impact,montoya2015stochastic,hobbs1993measuring,ozdemir2015economic,munoz2015multistage,dehghan2015reliability,moreira2016reliable}. The impact of wind power generation on the generation planning is presented in \cite{shortt2012accommodating}. A clustered unit commitment model is implemented in expansion planning problem and the effect of operational flexibility on the renewable generation planning is evaluated in \cite{palmintier2015impact}. A multi-stage expansion planning problem that considers the renewable energy generation and correlation is considered in \cite{montoya2015stochastic, hobbs1993measuring, ozdemir2015economic}. This approach is improved in \cite{munoz2015multistage} by including the topological changes. A tri-level robust expansion planning problem considering the uncertainty in renewable energy generation is presented in \cite{dehghan2015reliability}. A two stage co-optimization model for the expansion planning of renewable generation and transmission system under renewable uncertainty is presented in \cite{moreira2016reliable}. 

\par Expansion planning by considering both renewable generation and load growth has been a topic of interest \cite{li2017robust,alotaibi2018incentive,roldan2018robust,verastegui2019adaptive}. A robust coordinated generation and transmission expansion planning model considering the ramp uncertainty and net load output uncertainty with renewable energy integration is proposed in \cite{li2017robust}. A stochastic planning model to install distributed generation and feeders in the smart distribution system by probabilistically considering the uncertainty in system demand, and renewable generations is presented in \cite{alotaibi2018incentive}. A tri-level min-max-min approach which considers generation and transmission expansion planning in the upper level, realization of uncertainty in demand and renewable generation in the middle level, and operation in the lower level is presented in \cite{roldan2018robust}. An adaptive two-stage robust optimization model for generation and transmission expansion planning under daily operational uncertainty of load and renewable generation is presented in \cite{verastegui2019adaptive}.

\par Apart from expansion planning to match the growing demand and renewable integration of a power system, some researchers have also addressed this problem under considering environmental conditions. The need for enhancing the resilience of electricity grid against natural disasters has been extensively acknowledged \cite{executive2013economic, national2016analytic, wang2015research, bie2017battling}. A review of vulnerabilities in an electricity grid due to natural disasters can be found in \cite{waseem2020electricity}. An integrated electricity-natural gas system expansion planning model to enhance the resilience of electricity grid against natural disasters is presented in \cite{shao2017integrated}. A generation expansion planning considering the discrete climate change scenarios is presented in \cite{li2016stochastic}. A grid infrastructure planning to adapt to the changes in weather is presented in \cite{brockway2020weathering}. A two-stage robust optimization and co-expansion planning problem in integrated gas and power systems to increase the resilience against natural disasters is presented in \cite{zou2020resilient}. A generation and transmission expansion planning to increase the resilience of electricity grid against earthquake and floods is presented in \cite{hamidpour2021multi}. A security-constrained transmission expansion planning model by considering the impact of wind and natural disasters is presented in \cite{zhou2019security} but it does not consider the temporal and spatial characteristics of natural disasters.

\par 
A gap in expansion planning of electricity grid under severe wildfire conditions by taking into account the wildfire ignition score of power lines and mitigate the impact of wildfires by solar energy integration, modification of existing lines, and addition of new lines exists in the literature. This paper aims to fill this gap by considering a 10-year expansion planning problem in which the wildfire ignition risk of each power line is integrated. Three decision variables including solar energy integration, modification of existing lines, and addition of new lines are incorporated into the model to serve the load demand during wildfires. The uncertainty in wildfire ignition risk, load demand, and solar power generation are taken into consideration to render the real system operation.

\subsection{Summary of the Contributions}
 The main contributions of this paper are outlined as follows:
 \begin{enumerate}
 \item A 10-year expansion planning problem is formulated as a two-stage robust optimization problem. The first stage minimizes the operation cost by considering distributed solar generation, modification of power lines, and addition of new lines while the second stage realizes the uncertainty scenarios in system demand and solar energy generation. 
 \item The impact of deviation in wildfire ignition risk, system demand, and solar power generation 
 on the operation cost is assessed.
 \item The number of hours each line is energized and load shedding during a 10-year expansion planning scenario is evaluated.
 \item The impact of distributed solar generation, modification of existing lines, and addition of new lines to serve the affected customers due to severe wildfire weather conditions on the operation cost is assessed.
\end{enumerate}
The remainder of the paper is organized as follows. Section \ref{pblm_form_sol} describes the two-stage robust optimization problem formulation and its solution using the Column-and-Constraint Generation Algorithm (CCGA). Section \ref{case_study} describes the case studies and provides numerical results, while section \ref{conclusion} concludes and summarizes the paper.

\section{Problem Formulation and Solution Method} \label{pblm_form_sol}
The expansion planning problem is formulated as a two-stage robust optimization problem which is solved using CCGA to handle the uncertainties . The first-stage minimizes the operation cost by considering distributed solar generation, modification of existing power lines, and addition of new power lines. The \emph{here-and-now} decision variables $(I_{l,y}^{exs}, I_{l,y}^{mod}, \overline{P}_{y}^{R,i})$  are retrieved by solving the first-stage problem and then passed to the second-stage problem. The second-stage problem is \emph{wait-and-see} and realizes the uncertainty levels in system demand and solar energy generation. The steps involved in solving a two-stage robust optimization problem using CCGA are given as follows.

\begin{enumerate}[label=(\alph*)]
    \item Set iteration number $\varnothing=0$, lower bound ($\text{LB})=-\infty$, and upper bound ($\text{UB})=+\infty$. Risk tolerance level $\epsilon$ is selected by the system operator. The goal is to find the realization of uncertainty levels in demand and solar generation, based on the expansion planning decisions.
    \item Solve the first-stage problem, i.e. minimizing the operation cost by considering the solar installed capacity, modification of lines, and addition of new lines to serve the load during wildfire ignition scenario, as described in \eqref{fs}. 
    \begin{subequations}\label{fs}
    \begin{alignat}{2}
    &\underset{{P_{s,y}^{G,g},P_{s,y}^{D,i}}}{\textbf{min}} \sum\limits_{y \in \mathcal{Y}}\begin{Bmatrix}\sum\limits_{l\in \mathcal{L}} CN_y^l(I_{l,y}^{exs}-I_{l,y-1}^{exs})  \\
    & \hspace{-3.9cm}+\sum\limits_{i \in \mathcal{I}}CD_y(\overline{P}_{y}^{R,i}-\overline{P}_{y-1}^{R,i}) \\
    & \hspace{-3.8cm}+\sum\limits_{l\in \mathcal{L}}CM_y^l(I_{l,y}^{mod}-I_{l,y-1}^{mod})\end{Bmatrix}+e\label{pblm_1a}\\
    & \textbf{subject to:} \nonumber\\
    &e \geq \sum\limits_{y\in \mathcal{Y}}\sum\limits_{s\in \mathcal{S}}\begin{Bmatrix} R_{s} \{\sum\limits_{g\in \mathcal{G}}\sum\limits_{z\in \mathcal{Z}} c_g^{z} P_{s, y}^{g,z (\varnothing)} \\
    & \hspace{-3.6cm}+\sum\limits_{i\in \mathcal{I}} K (P_{s,y}^{D,i(\varnothing)}-P_{s,y}^{d,i(\varnothing)})\}\end{Bmatrix}\label{pblm_1b} \\
    & \sum\limits_{z\in \mathcal{Z}} P_{s,y}^{g, z(\varnothing)} = P_{s,y}^{G,g(\varnothing)},\hspace{0.2cm}  \forall g \in \mathcal{G}, s \in \mathcal{S},  y \in \mathcal{Y} \nonumber \\
    & \hspace{6.1cm} :\lambda1_{s,y}^g\label{pblm_1c}\\
    & 0 \leq P_{s,y}^{g,z(\varnothing)} \leq \overline P_s^{g,z}, \hspace{0.2cm} \forall g \in \mathcal{G}, s \in \mathcal{S}, y \in \mathcal{Y}, z \in \mathcal{Z} \nonumber\\
    & \hspace{6.1cm} :\overline{\mu1}_{s,y}^{g,z} \label{pblm_1d}\\
    & \underline{P}_{s}^g \leq P_{s,y}^{G,g(\varnothing)} \leq \overline{P}_{s}^g, \hspace{0.2cm} \forall g \in \mathcal{G}, s \in \mathcal{S}, y \in \mathcal{Y} \hspace{0.2cm} :\underline{\mu2}_{s,y}^g, \nonumber \\ &\hspace{6.4cm}\overline{\mu2}_{s,y}^g \label{pblm_1f}\\
    & 0 \leq P_{s,y}^{R,i(\varnothing)} \leq \overline{P}_{y}^{R,i}\kappa_{s,y}^{i(\varnothing)}, \hspace{0.2cm} \forall i \in \mathcal{I}, s \in \mathcal{S}, y \in \mathcal{Y}\nonumber\\
    & \hspace{5.2cm} :\underline{\mu3}_{s,y}^{i}, \overline{\mu3}_{s,y}^{i} \label{pblm_1g}\\
    & 0 \leq P_{s,y}^{d,i(\varnothing)} \leq \overline{ P}_{s,y}^{D,i(\varnothing)},\hspace{0.2cm} \forall i \in \mathcal{I}, s \in \mathcal{S}, y \in \mathcal{Y} \hspace{0.2cm} :\overline{\mu4}_{i}^{s,y} \label{pblm_1h}\\
    &  P_{s, y}^{d,i(\varnothing)} + \sum\limits_{l \in \mathcal{L}^{fr}_{i}}P_{s, y}^{L,l(\varnothing)} = \sum\limits_{l \in \mathcal{L}^{to}_{i}}P_{s, y}^{L,l(\varnothing)} +\sum\limits_{g \in G_i}P_{s, y}^{G,g(\varnothing)}\nonumber\\
    &+\sum\limits_{R \in R_i} P_{s, y}^{R, i(\varnothing)},\hspace{0.2cm} \forall i \in \mathcal{I}, s \in \mathcal{S}, y \in \mathcal{Y} \hspace{0.4cm} :\lambda2_{s,y}^i\label{pblm_1e}\\
    & -M(2-I_{s,y}^l-I_{l,y}^{exs})+P_{s,y}^{L,l(\varnothing)} \leq \nonumber\\
    & \frac{\sum\limits_{i\in B_{enter.}^l}\theta_{s,y}^{i(\varnothing)} - \sum\limits_{i\in B_{leav.}^l}\theta_{s,y}^{i(\varnothing)}}{x_l} \leq  P_{s,y}^{L,l(\varnothing)}+M(2-I_{s,y}^l \nonumber\\
    & -I_{l,y}^{exs}), \hspace{0.2cm} \forall l \in \mathcal{L}, s \in \mathcal{S}, y \in \mathcal{Y} \hspace{0.7cm}:\underline{\mu5}_{s,y}^{l}, \overline{\mu5}_{s,y}^{l} \label{pblm_1i}\\
    & -\overline{P}_{s}^{l} I_{s,y}^l \leq P_{s,y}^{L,l(\varnothing)} \leq \overline{P}_s^l I_{s,y}^l, \hspace{0.2cm}  \forall l \in \mathcal{L}, s \in \mathcal{S}, y \in \mathcal{Y} \nonumber\\
    & \hspace{5.3cm} :\underline{\mu6}_{s,y}^l, \overline{\mu6}_{s,y}^l \label{pblm_1j}\\
    & SC_{s,y}^{l(\varnothing)} \geq (\psi_{s,y}^{l} + \bigtriangleup \psi_{s,y}^{l})  (I_{s,y}^l -\delta aux_{s,y}^{l}),\hspace{0.2cm} \forall l \in \mathcal{L}, \nonumber\\ 
    & \hspace{4.0cm} s \in \mathcal{S}, y \in \mathcal{Y} \hspace{0.2cm} :\underline{\mu7}_{s,y}^{l} \label{pblm_1j_1}\\
    & aux_{s,y}^{l} \leq I_{l,y}^{mod}, \hspace{1.7cm} \forall l \in \mathcal{L}, s \in \mathcal{S},  y \in \mathcal{Y} \label{pblm_1j_2}\\
    & aux_{s,y}^{l} \leq I_{s,y}^{l}, \hspace{1.9cm} \forall l \in \mathcal{L}, s \in \mathcal{S},  y \in \mathcal{Y} \label{pblm_1j_3}\\
    & aux_{s,y}^{l} \geq I_{l,y}^{mod} + I_{s,y}^{l} -1, \hspace{0.1cm} \forall l \in \mathcal{L}, s \in \mathcal{S},  y \in \mathcal{Y}\label{pblm_1f_4}\\
    & \sum\limits_{l\in \mathcal{L}} SC_{s,y}^{l(\varnothing)} \leq \mathcal{E},\hspace{1.3cm} \forall s \in \mathcal{S}, y \in \mathcal{Y} \hspace{0.2cm} :\mu8_{s,y}  \label{pblm_1k}\\
    & I_{s,y}^{l} \leq I_{l,y}^{exs},  \hspace{2.4cm} \forall  l  \in \mathcal{L}, s \in \mathcal{S}, y \in \mathcal{Y} \label{pblm_1l_1} \\
    & I_{l,y}^{exs} \geq I_{l,y-1}^{exs}, \hspace{3.0cm} \forall l  \in \mathcal{L}, y \in \mathcal{Y} \label{pblm_1l_1_a} \\
    & I_{l,y}^{exs} \geq \rho^l, \hspace{3.6cm} \forall l \in \mathcal{L}, y \in \mathcal{Y} \label{flag}\\
    & P_{s,y}^{D,i(\varnothing)} \in \begin{bmatrix}P_{s,y}^{D,i,0}-\bigtriangleup P_{s,y}^{D,i},\hspace{0.1cm} P_{s,y}^{D,i,0}+\bigtriangleup P_{s,y}^{D,i} \end{bmatrix}, \nonumber\\ 
    & \hspace{4.2cm} \forall i \in \mathcal{I}, s \in \mathcal{S},  y \in \mathcal{Y}\label{pblm_ml}\\
    & \kappa_{s,y}^{i(\varnothing)} \in \begin{bmatrix}\kappa_{s,y}^{i,0}-\bigtriangleup \kappa_{s,y}^{i},\hspace{0.1cm}\kappa_{s,y}^{i,0}+\bigtriangleup \kappa_{s,y}^{i} \end{bmatrix}, \hspace{0.2cm} \forall i \in \mathcal{I}, s \in \mathcal{S}, \nonumber\\
    & \hspace{6.4cm} y \in \mathcal{Y} \label{pblm_1n}
    \end{alignat}
    \end{subequations}
    
The objective function minimizes the operation cost by considering the solar power generation, modification of existing power lines, and addition of new lines as given in \eqref{pblm_1a}. When a new line is installed, the expression ($I_{l,y}^{exs}-I_{l,y-1}^{exs}$) is 1 and $CN_y^l$ is added to the objective. $CN_y^l$ is zero for existing lines and it forces $I_{l,y}^{exs}$ to become 1 for existing lines. Whenever a line is modified, ($I_{l,y}^{mod}-I_{l,y-1}^{mod}$) becomes 1 and $CM_y^l$ participates in the objective. When solar installation of current year is greater than previous year, the difference ($\overline{P}_{y}^{R,i}-\overline{P}_{y-1}^{R,i}$) is multiplied with $CD_y$ and contributes to the objective. The operation cost of \emph{wait-and-see} which is the maximum value that can be procured is given by \eqref{pblm_1b}. It is the sum of power generation cost from generating units and penalty for not serving the demand.
The dispatched power of each generation unit is equal to the  summation of its segmented powers, as constrained by \eqref{pblm_1c}. The generation power of each segment is limited based on their capacities by \eqref{pblm_1d}. The power generation by each generator is constrained in \eqref{pblm_1f}. The solar energy generation is constrained in \eqref{pblm_1g} where $\kappa_{s,y}^{i}$ ranges between 0 to 1 and demand served is limited by \eqref{pblm_1h}. The nodal power flow balance at each bus is given in \eqref{pblm_1e}. The power flow equation is enforced by \eqref{pblm_1i}. Here, if a line exists in the network and is energized, these inequality constraints turn into an equality equation. The thermal limit of power flowing through energized lines are enforced in \eqref{pblm_1j}. The fire ignition score of a modified line is given by the constraint \eqref{pblm_1j_1}. For all energized lines, $SC$ should exceed the ignition score of that line, unless the line is modified. Here, the auxiliary variable $aux$ is used to determine if an energized line is modified, i.e. $aux_{s,y}^{l} = I_{l,y}^{mod} \cdot I_{s,y}^{l}$. However, to maintain linearity, this bi-linear relation is replaced with the equivalently linear constraints \eqref{pblm_1j_2}, \eqref{pblm_1j_3}, and \eqref{pblm_1f_4}. The aggregated quantified fire ignition score of power lines is limited by risk tolerance as given in \eqref{pblm_1k}. 
According to \eqref{pblm_1l_1}, once a line is physically present in the network, its existence status can not become 0. The energization/de-energization decision of each line is also meaningful only for physically present lines, which is forced by \eqref{pblm_1l_1_a}.
The uncertainty in system solar energy generation and demand for each scenario are calculated using \eqref{pblm_ml} and \eqref{pblm_1n}. 
\par 
 The quantified fire ignition score $\psi$ in \eqref{pblm_1j_1} is obtained from our previous work \cite{waseem_1}. A visualization of diverse wind speeds in a 6-bus test case is shown in Fig. \ref{6_bus_demo}. The fire ignition score is calculated by physically modeling the 3D non-linear vibrational motion of power lines by applying the Hamilton's principle. The 3D equations are simplified to 2D equations by applying the boundary and modeling assumptions. These 2D continuous partial differential equations (PDEs) are reformed into discrete PDEs using Galerkin method. The resulting equations represent the in-plane and out-of-plane vibrational motion of power lines and are solved using Runge-Kutta method by considering numerous physical, structural, and meteorological features including span of power line, phase clearance, conductor diameter, wind speed, wind gust, and wind direction. A surrogate machine learning model is developed to forecast the fire ignition score. A sample of fire ignition score based on various features is shown in Table \ref{samples_dataset}.
    \item By solving the minimization problem \eqref{fs}, $\text{LB}$ is set as its objective value and binary decision variables $I_{l,y}^{exs}$, $I_{l,y}^{mod}$, $I_{s,y}^l$, and $aux_{s,y}^l$ are obtained.
    \item Solve the second stage problem to find the new realizations of uncertain binary decision variables $u_{s,y}^{D,i}$, $v_{s,y}^{D,i}$, $u_{s,y}^{R,i}$, and $v_{s,y}^{R,i}$.
    
    The dual formulation of \eqref{fs} is given in \eqref{dual}. 
    \begin{subequations}\label{dual} 
    \begin{alignat}{2}
    &\underset{{P_{s,y}^{D,i},\overline{P}_{y}^{R,i}}}{\textbf{max}}\sum\limits_{s\in\mathcal{S}}\sum\limits_{y\in\mathcal{Y}} \nonumber\\
    &\begin{Bmatrix}\sum\limits_{g\in \mathcal{G}}\sum\limits_{z\in \mathcal{Z}} -\overline{P}_s^{g,z}\overline{\mu1}_{s,y}^{g,z}+\sum\limits_{g\in \mathcal{G}}(\underline{P}_s^g\underline{\mu2}_{s,y}^{g}-\overline{P}_s^g\overline{\mu2}_{s,y}^{g})\\
    &\hspace{-9.8cm}- \sum\limits_{l\in \mathcal{L}} \{\overline{P}_{s}^l (\hat{I}_{s,y}^l\underline{\mu6}_{s,y}^{l}+\hat{I}_{s,y}^l\overline{\mu6}_{s,y}^{l})\\
    &\hspace{-9.3cm}+M(2-\hat{I}_{s,y}^l-\hat{I}_{l,y}^{exs})(\underline{\mu5}_{s,y}^{l}+\overline{\mu5}_{s,y}^{l}) \\
    &\hspace{-8.8cm}- (\psi_{s,y}^{l} + \bigtriangleup \psi_{s,y}^{l})(\hat{I}_{s,y}^l-\delta\hat{aux}_{s,y}^l)\underline{\mu7}_{s,y}^{l}\}\\
    &\hspace{-7.6cm}-\sum\limits_{i\in \mathcal{I}}\hat{\overline{P}}_{y}^{R,i}(\kappa_{s,y}^i+\triangle\kappa_{s,y}^i u_{s,y}^{R,i}-\triangle\kappa_{s,y}^i v_{s,y}^{R,i})\overline{\mu3}_{s,y}^i\\
    &\hspace{-7.4cm}-\sum\limits_{i\in \mathcal{I}}\overline{P}_{s,y}^{D,i}({P}_{s,y}^{d,i}+\triangle{P}_{s,y}^{d,i} u_{s,y}^{D,i}-\triangle{P}_{s,y}^{d,i} v_{s,y}^{D,i})\overline{\mu4}_i^{s,y} \\
    \hspace{-5.8cm} -\mathcal{E} \mu8_{s,y}\end{Bmatrix}
    \label{dual_1a} \\
    & \textbf{subject to:} \nonumber\\
    & -\lambda1_{s,y}^g - \sum\limits_{i\in B_{g}}\lambda2_{s,y}^i + \underline{\mu2}_{s,y}^g - \overline{\mu2}_{s,y}^g=0, \hspace{0.2cm} \forall g \in \mathcal{G}, \nonumber\\
    & \hspace{4.0cm} s \in \mathcal{S}, y \in \mathcal{Y} \hspace{0.2cm} :P_{s,y}^{G,g} \label{dual_1b} \\
    & -\sum\limits_{i\in B_{enter.}^l}\lambda2_{s,y}^i+\sum\limits_{i\in B_{leav.}^l}\lambda2_{s,y}^i -\underline{\mu5}_{s,y}^l+ \overline{\mu5}_{s,y}^l \nonumber\\
    &  
     + \underline{\mu6}_{s,y}^l - \overline{\mu6}_{s,y}^l = 0, \hspace{0.0cm}  \forall l \in \mathcal{L}, s \in \mathcal{S}, y \in \mathcal{Y},  \hspace{0.0cm}:P_{s,y}^{L,l}\label{dual_flow_line}\\
    & \lambda2_{s,y}^i - \overline{\mu4}_{i}^{s,y} \leq -KR_s, \hspace{0.2cm}  \forall i \in \mathcal{I}, s \in \mathcal{S}, y \in \mathcal{Y}  \nonumber \\
    & \hspace{6.3cm} :P_{s,y}^{d,i}\label{dual_1d} \\
    & -\lambda2_{s,y}^i - \overline{\mu3}_{s,y}^{i} \leq 0,  \hspace{0.0cm}  \forall i \in \mathcal{I}, s \in \mathcal{S}, y \in \mathcal{Y} \hspace{0.0cm} :P_{s,y}^{R,i}\label{dual_1e} \\
    & \lambda1_{s,y}^g - \overline{\mu1}_{s,y}^{g,z} \leq c_g^{z}R_s,  \hspace{0.2cm}  \forall g \in \mathcal{G}, s \in \mathcal{S}, y \in \mathcal{Y}, z \in \mathcal{Z} \nonumber \\
    & \hspace{6.3cm} :P_{s, y}^{g,z}\label{dual_1f} \\
    &\sum\limits_{l\in L^{fr}_{i}}\frac{\underline{\mu5}_{s,y}^l-\overline{\mu5}_{s,y}^l}{x_l}-\sum\limits_{l\in L^{to}_{i}} \frac{\underline{\mu5}_{s,y}^l-\overline{\mu5}_{s,y}^l}{x_l}=0, \nonumber \\
    & \hspace{3.1cm} \forall i \in \mathcal{I}, s \in \mathcal{S}, y \in \mathcal{Y} \hspace{0.2cm}:\theta_{s,y}^i\label{dual_1g} \\
    & \underline{\mu7}_{s,y}^{l}-\mu8_{s,y} \leq 0, \hspace{0.2cm} \forall l \in \mathcal{L}, s \in \mathcal{S}, y \in \mathcal{Y} \hspace{0.0cm} :SC_{s,y}^l\label{dual_1h}\\
    & \sum\limits_{i\in \mathcal{I}}\sum\limits_{s\in \mathcal{S}}\sum\limits_{y\in \mathcal{Y}} (u_{s,y}^{R,i} + u_{s,y}^{D,i}) \leq E \label{dual_1i}\\
    & \overline{\mu}, \underline{\mu} \geq 0, \lambda \label{dual_1j}
    \end{alignat}
    \end{subequations}

    The dual objective is given in \eqref{dual_1a} and dual constraints are given in \eqref{dual_1b}-\eqref{dual_1j}. The dual of generation dispatch of generator decision variable is given in \eqref{dual_1b}. The dual of power flow of a line continuous decision variable is given in \eqref{dual_flow_line}. The dual of schedule demand decision variable is given in \eqref{dual_1d}. The dual of solar energy dispatch decision variable is given in \eqref{dual_1e}. The dual of generation from segment $g$ of cost curve decision variable is given in \eqref{dual_1f}. The dual of bus voltage angle decision variable is given in \eqref{dual_1g}. The dual of conductor clashing score decision variable is given in \eqref{dual_1h}. Sum of uncertainty levels in demand and solar generation is less than the budget of uncertainty as given in \eqref{dual_1i}. $\lambda$ is a free variable in \eqref{dual_1j}. The hat ( $\hat{}$ ) on \emph{here-and-now} decision variables in the dual objective denotes these variables are passed from first stage to the second stage problem.
    \par It is noticed than in the objective function of the dual problem, some non-linear terms appear that contain binary-to-continuous variable multiplication.
    To avoid solving a non-linear problem of this kind, a linearization process is applied which is exemplified in \eqref{dual_linear}. Here, the non-linear term in \eqref{dual_linear_aa} is linearized by \eqref{dual_linear_bb}-\eqref{dual_linear_dd}, where $v_{s,y}^{R,i}$ is a binary variable. Two auxiliary positive continuous variables $\nu$ and $\omega$ which are bounded between 0 and M are required for this linearization step.
    \begin{subequations}\label{dual_linear}
    \begin{alignat}{2}
    & \nu_y^s = v_{s,y}^{R,i}\overline{\mu}_{y}^{s},\hspace{0.1cm} v_{s,y}^{R,i} \in \begin{Bmatrix}0,1\end{Bmatrix} \hspace{0.1cm} \forall i \in \mathcal{I}, s \in \mathcal{S}, y \in \mathcal{Y} \label{dual_linear_aa} \\
    & \nu_y^s = \overline{\mu}_{y}^{s} - \omega_y^s, \hspace{0.2cm} \forall s \in \mathcal{S}, y \in \mathcal{Y} \label{dual_linear_bb} \\
    & 0 \leq \nu_y^s \leq M\cdot v_{s,y}^{R,i}, \hspace{0.2cm} \forall i \in \mathcal{I}, s \in \mathcal{S}, y \in \mathcal{Y} \label{dual_linear_cc} \\
    & 0 \leq \omega_y^s \leq M\cdot (1-v_{s,y}^{R,i}), \hspace{0.2cm} \forall i \in \mathcal{I}, s \in \mathcal{S}, y \in \mathcal{Y} \label{dual_linear_dd}
    \end{alignat}
    \end{subequations}
    \item Upon solving the second-stage problem \eqref{dual}, new realization of uncertain variables are obtained based on the binary decision variables calculated in \eqref{fs}. The value of $\text{UB}$ is set equal to the objective value of \eqref{dual}, and we check if $ \text{UB}-\text{LB} \leq \epsilon$ holds. If this condition is satisfied, the algorithm is converged as the primal and dual problems return similar objective values, and the solution process is terminated. Otherwise, a new realization of uncertainty levels is obtained by adding \eqref{change_1} and \eqref{change_2} to the first stage problem \eqref{fs}.
    \item Increment the iteration $\varnothing=\varnothing+1$ and return to step (b).
    
\begin{figure}[bht] 
\centering
   \includegraphics[width=0.4\textwidth]{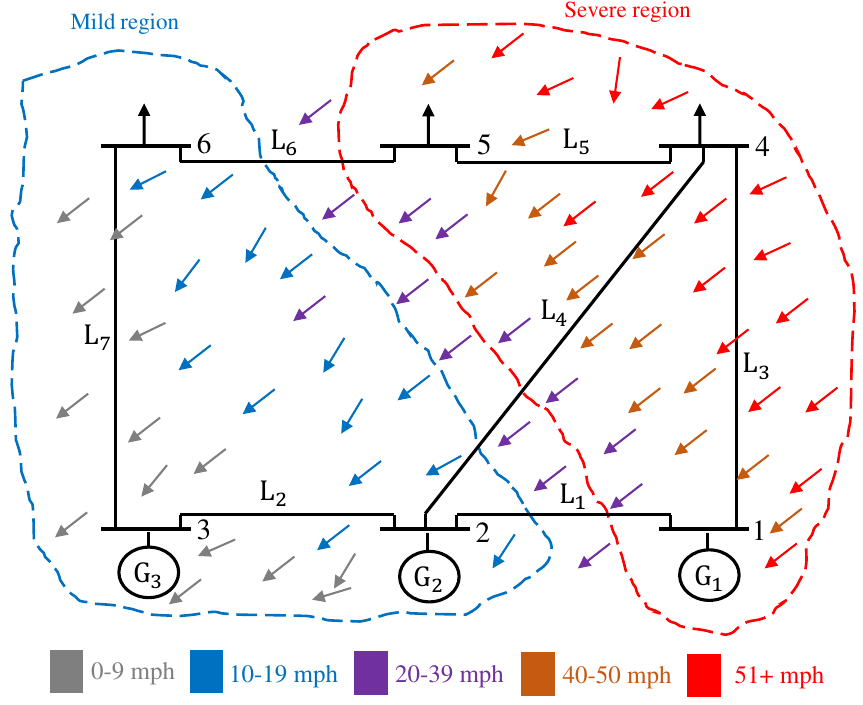}
\caption{Divergent wind speeds results in different fire ignition scores in a 6-bus test case }
\label{6_bus_demo} 
\end{figure}

\end{enumerate}
{\setcounter{table}{0}
\begin{table}[h!]
    \vspace{-0.31cm}
	\small \centering
	\caption{\color{black}A Sample from Dataset in which Fire Ignition Scores are Predicted by Varying Different Features} 
	\vspace{-0.2cm}
	\begin{tabular}{M{0.4cm}M{1.27cm}M{0.49cm}M{0.70cm} M{1.29cm}M{1.21cm}M{0.54cm}} \hline\hline
	    Span (ft)  & Conductor diameter (mm) & Wind speed (mph) & Wind gust (mph) & Phase clearance (ft) & Wind direction (\textdegree) & Score [0,1)  \\ \hline 
		600& 33.03& 22& 27& 0.5&45  &0  \\
	    800& 34.02& 40& 36&0.5& 180   &0.01\\
		500& 33.03& 40& 45&0.7& 45 &  0.08\\
		1000& 31.05& 63& 67&0.5& 90  &0.12 \\
		400& 33.03& 58& 31&0.9&315 & 0.17 \\
		300& 33.03& 67& 67&0.5& 315   &0.45\\\hline\hline
	\end{tabular}
	\label{samples_dataset}
\end{table}}

Once the condition in step (e) of CCGA is not satisfied, the solution of the second-stage problem is used to create additional constraints which are then augmented to the first stage problem. Using the solution of \eqref{dual}, the new realization of the uncertain variables is procured by \eqref{change}, where $\triangle$ in \eqref{change} denotes the magnitude of uncertainty ($\pm10\%$, $\pm20\%$, \text{etc.}).

\begin{subequations}\label{change}
\begin{alignat}{2}
& P_{s,y}^{D,i(\varnothing+1)} = P_{s,y}^{D,i(\varnothing)} +  \triangle P_{s,y}^{d,i}\cdot u_{s,y}^{D,i(\varnothing)}  - \triangle P_{s,y}^{d,i}\cdot v_{s,y}^{D,i(\varnothing)}, \nonumber\\
& \hspace{4.9cm} \forall i \in \mathcal{I}, s \in \mathcal{S}, y \in \mathcal{Y}\label{change_1} \\
& \kappa_{s,y}^{i(\varnothing+1)} = \kappa_{s,y}^{i(\varnothing)} + \triangle \kappa_{s,y}^i\cdot u_{s,y}^{R,i(\varnothing)} - \triangle \kappa_{s,y}^i \cdot v_{s,y}^{R,i(\varnothing)}, \nonumber\\
& \hspace{4.9cm} \forall i \in \mathcal{I}, s \in \mathcal{S}, y \in \mathcal{Y}\label{change_2} 
\end{alignat}
\end{subequations}

\begin{figure}[h!] 
    \centering
       \includegraphics[width=0.45\textwidth]{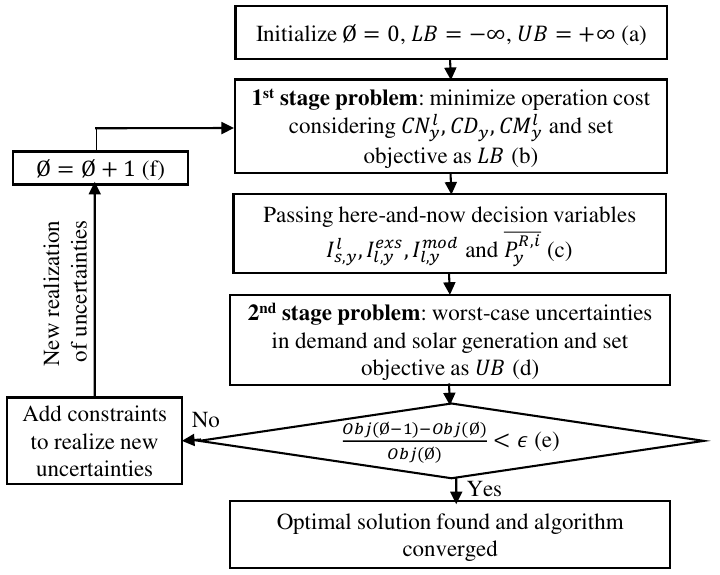}
    \caption{Flow diagram of Column-and-Constraint Generation Algorithm}
    \label{ccga} 
\end{figure}

The flow diagram of CCGA is shown in Fig. \ref{ccga}. Initially, the first stage objective is set to negative infinity and second stage objective is set to positive infinity. In the first stage problem, the operation cost is minimized considering the cost of solar installation, modification of lines, and addition of new lines. LB is updated based on the current objective of the first stage problem. The here-and-now variables including decision for lines energization status, installation and modification of lines, and solar capacity installed are passed to the second stage problem. The second stage problem realizes the uncertainty levels in demand and solar generation and updates UB objective. The convergence of the LB and UB objectives is evaluated by checking the difference between the current and previous iterations and calculating if it is less than a threshold. If both LB and UB are less than the threshold, then the algorithm converges; otherwise, new realizations of uncertainty levels are accomplished by equations \eqref{change_1} and \eqref{change_2}. When the first stage decision variables are not robustly feasible, the dispatch is not possible and the second stage returns only the worst-case realizations of uncertainty levels $u^{\varnothing+1}$.
However, if the first stage decision variables are robustly feasible, the worst-case operation cost is returned by the second stage and the corresponding scenario of worst-case uncertainty $u^{\varnothing+1}$.

\par
The uncertainty in system demand and solar energy generation is limited by the budget of uncertainty. It confines the binary variables that render the worst-case realization of uncertain variables. It is selected by the system operator and represents the magnitude of uncertainty period. The aim of second stage uncertainty variables is to minimize the operation cost of first stage and second stage based on the budget of uncertainty.

\section{Case Study}\label{case_study}
In this section, two case studies are presented to evaluate the effectiveness of the presented two-stage robust optimization problem for determining the decision variables in a 10-year expansion planning scenario considering wildfire ignition risk. The first case study analyses a 6-bus system and the second one uses IEEE 118-bus system. The parameter $\delta$ is set to 0.5 to consider that the fire ignition score is reduced by half once a line is  modified. A personal computer with 3.60 GHz processor and 16.0 GB RAM is employed. CPLEX 12.9 is used for 6-bus system simulations and Gurobi 9.1 is used for 118-bus system simulations to perform these studies.
\subsection{6-Bus Power System}
\begin{figure}[bht] 
    \centering
       \includegraphics[width=0.45\textwidth]{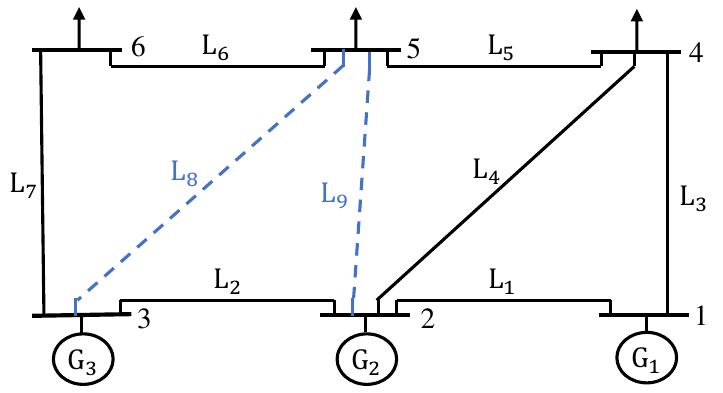}
    \caption{A test case of 6-bus system}
    \label{6_bus} 
\end{figure}

In this part, a 6-bus power system network is considered as shown in Fig. \ref{6_bus}. It consists of 3 generating units and 7 power lines with two additional candidate lines ($L_8$ and $L_9$). The characteristics of generation units and power lines are given in Tables \ref{generation_unit_1} and \ref{transmission_line_}, respectively.

\begin{table}[bht]
	\vspace{-0.31cm}
	\small \centering
	\caption{\color{black}Generation Unit Characteristics in the 6-Bus Test Case} 
	\vspace{-0.2cm}
	\begin{tabular}{M{0.6cm} M{0.8cm} M{0.8cm} M{0.6cm} M{0.8cm} M{1.0cm}} \hline\hline 
		Unit& P$_{\text{min}}$ (MW) & P$_{\text{max}}$ (MW)&  \textbf{a} $(\$)$&\textbf{b} $(\$$/MW) & c $(\$$/MW$^2$)   \\\hline
        G$_1$  & 100& 220 &177 &13.5 & 0.00045\\
	    G$_2$  &10 &100 &130 &40 & 0.001\\
	    G$_3$  &10 &40  & 137&17.7 &0.005\\\hline\hline
	\end{tabular}
	\vspace{-0.21cm}
	\label{generation_unit_1}
\end{table}

Two scenarios namely low wind speed and high wind speed are considered. The low wind speed scenario has less fire ignition score while the high wind speed scenario has high fire ignition score. The total fire ignition score, net demand, and average wind speed for 10-year period in the 6-bus test case is shown in Table \ref{overall_fire_ignition_score_10_year}. The total fire ignition score for each year is contributed by the energized lines based on their physical existence in the low/high wind speed regions. 

\begin{table}[h!]
	\vspace{-0.31cm}
	\small \centering
	\caption{\color{black}Power line Characteristics in the 6-Bus Test Case} 
	\vspace{-0.2cm}
	\begin{tabular}{M{0.1cm} M{0.4cm} M{0.3cm} M{0.8cm} M{1.6cm}M{1.37cm}M{1.37cm}} \hline \hline 
		ID& From Bus & To Bus & Reactance (p.u.)& Maximum Rating (MW) & $\psi_{1-6,1}^{l}$ for low wind speed & $\psi_{1-6,1}^{l}$ for high wind speed \\ \hline
        L$_1$  & 1 & 2  &0.2 &200 & 0.0743 & 0.6330\\
	    L$_2$  &2 &3 &0.25 &100 & 0.0375 & 0.6432\\
	    L$_3$  &1 &4 &0.2 & 100 & 0.0251 & 0.6483\\
	    L$_4$  & 2& 4  & 0.1& 100 & 0.0189 & 0.6534\\
	    L$_5$  &4 &5 & 0.4& 100 & 0.0152 & 0.6584\\
	    L$_6$  &5 &6 &0.3 &100 & 0.0127 & 0.6636\\
	    L$_7$  &3 &6 &0.1 &100 & 0.0109 & 0.6687\\
	    $\bm{L_8}$ &\bf{3} &\bf{5} &\bf{0.26} &\bf{100} & \bf{0.0096 }& \bf{0.6738}\\
	    $\bm{L_9}$  &\bf{2} &\bf{5} &\bf{0.3} &\bf{100} & \bf{0.0086} & \bf{0.6789} \\\hline\hline
	\end{tabular}
	\label{transmission_line_}
\end{table}

\begin{table}[h!]
	\vspace{-0.31cm}
	\small \centering
	\caption{\color{black}Total Fire Ignition Score, Net Demand, and Average Wind Speed for 10-year Period in the 6-bus Test Case} 
	\vspace{-0.2cm}
	\begin{tabular}{M{0.15cm} M{1.7cm} M{1.38cm} M{1.02cm} M{0.9cm} M{0.95cm}}\hline\hline 
	Years & $\psi_{1-6,y}^{1-9}$ for low wind speed & $\psi_{1-6,y}^{1-9}$ for high wind speed & Net demand (MW) & Average of low wind speed (mph)& Average of high wind speed (mph)\\ \hline
     1 & 0.21 & 5.92 & 1078 & 37 & 62\\
	 2 & 0.22 & 6.38 & 1099 & 38 & 64\\
	 3 & 0.23 & 6.83 & 1121 & 41 & 65\\
	 4 & 0.24 & 7.29 & 1144 & 42 & 66\\
	 5 & 0.25 & 7.75 & 1167 & 43 & 68\\
	 6 & 0.20 & 5.37 & 1190 & 34 & 60\\
	 7 & 0.19 & 5.09 & 1214 & 32 & 59\\
	 8 & 0.18 & 4.83 & 1238 & 30 & 58\\
	 9 & 0.17 & 4.57 & 1263 & 29 & 56\\
	 10 & 0.16 & 4.49 & 1288 & 28 & 55\\\hline\hline
	\end{tabular}
	\vspace{-0.21cm}
	\label{overall_fire_ignition_score_10_year}
\end{table}

\subsubsection{Impact of Weather Conditions with Fire Hazard on Energization of Power Lines}
\begin{figure}[bht] 
    \centering
       \includegraphics[width=0.48\textwidth]{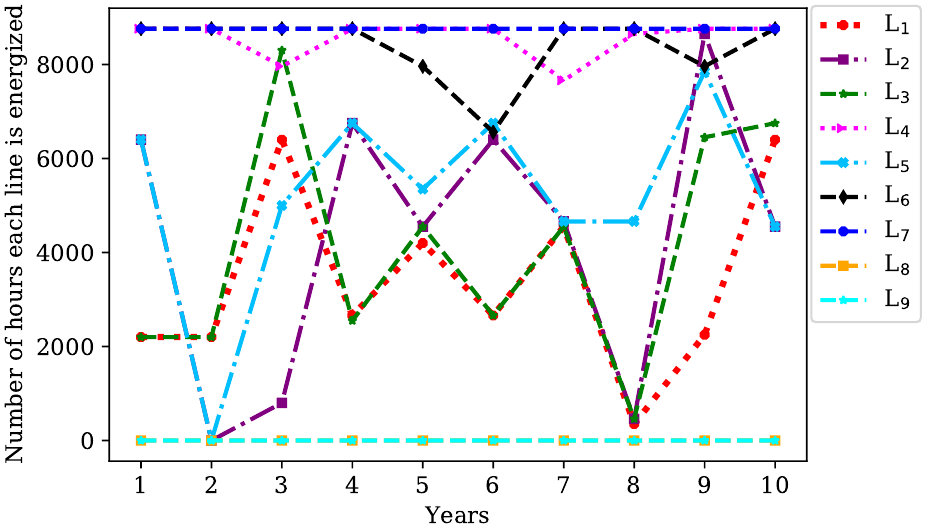}
    \caption{Number of hours lines are energized during the low wind speed scenario for a 10-year period in the 6-bus test case}
    \label{lines_low} 
\end{figure}

In this part, the impacts of severe weather conditions on the energization of power lines during a 10-year period for the low wind speed and high wind speed scenarios are assessed. During the low wind speed scenario, the energization status of lines is shown in Fig. \ref{lines_low}. It depicts how many hours each line is energized out of 8760 hours during a year for total of 10 years. 
Is is noticed that in the low wind speed scenario, line $L_7$ is energized at all times throughout the 10-year period. This is because this line is located in a region with comparatively lower wind speed than other regions in the 6 bus network, and it is also connected to high demands. The energization period of all other lines $L_1$ - $L_6$ is varying throughout the 10-year period. Noticeably, $L_1$ is energized for the smallest duration throughout this period. The reason is that this line has higher conductor clashing score among other lines. The variation in energization period of lines can be associated with the variation in conductor clashing score and load demand. 
In this scenario, the candidate lines $L_8$ and $L_9$ remain uninstalled and thus de-energized during the 10-year period because the original system with lines ($L_1$ - $L_7$) has sufficient capacity to supply the growing load demand over the 10-year horizon. 

\par
\begin{figure}[bht] 
    \centering
       \includegraphics[width=0.48\textwidth]{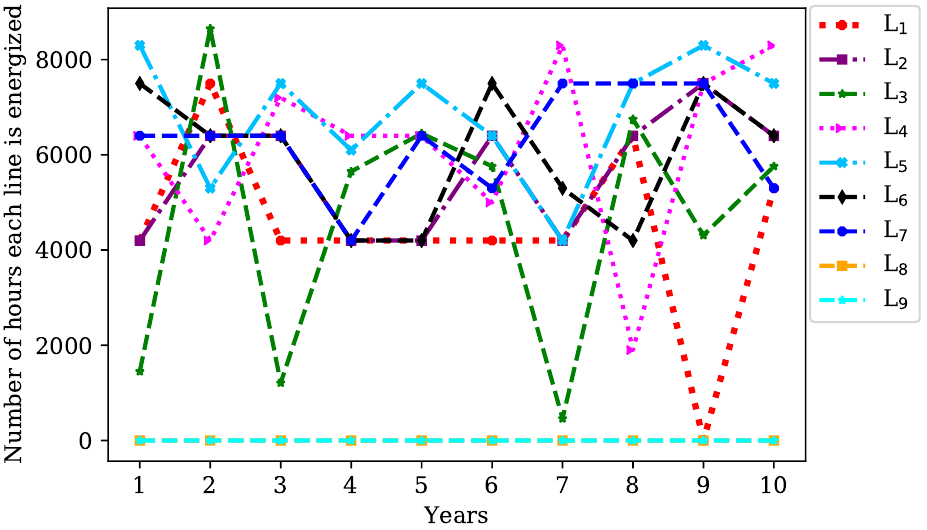}
    \caption{Number of hours lines are energized during the high wind speed scenario for a 10-year period in the 6-bus test case}
    \label{lines_high} 
\end{figure}

The energization status of power lines during the high wind speed scenario is shown in Fig. \ref{lines_high}. Higher wind speeds are more likely to cause conductor clashing, thereby they lead to more frequent line de-energization. Compared with the low wind speed scenario where line $L_7$ was energized 100\% of the time, this time this line is energized only around 69\% of the 10-year simulation period. Similarly, all of the other lines $L_1$ - $L_6$ are also faced to decrease their duration of energized hours. At years 3 and 7, Lines $L_1$ and $L_7$ are both fully de-energized due to the high wind speed scenario in these years which lead to higher conductor clashing scores.
In this case, the decision is that no candidate line needs to be installed. Due to their higher conductor clashing score, the candidate lines are not energized to help primary lines serve the load .While it is costly to add new lines to the system, the results suggest that installation of more solar generation units is beneficial to the system's operation in the high wind speed scenario as shown in Table \ref{total_solar_10_year}.

\begin{table}[h!]
	\vspace{-0.31cm}
	\small \centering
	\caption{\color{black}Total Solar Capacity Installed (MW) for 10-year Period in the 6-Bus Test Case} 
	\vspace{-0.2cm}
	\begin{tabular}{M{1.6cm} M{0.2cm} M{0.2cm} M{0.2cm} M{0.2cm}M{0.2cm} M{0.2cm} M{0.2cm} M{0.2cm} M{0.2cm}M{0.35cm}}\hline\hline 
	Years & 1 & 2 & 3 & 4 & 5 & 6 & 7 & 8 & 9 & 10 \\ \hline
    Low wind scenario  & 57 & 57 & 57 & 57 & 57 & 57 & 57 & 57 & 57 & 57\\
	High wind scenario & 276 & 276 & 276 & 276 & 276 & 276 & 276 & 276 & 276 & 276\\\hline\hline
	\end{tabular}
	\vspace{-0.21cm}
	\label{total_solar_10_year}
\end{table}

\subsubsection{Installed Solar Capacity}
\begin{figure}[bht] 
    \centering
       \includegraphics[width=0.48\textwidth]{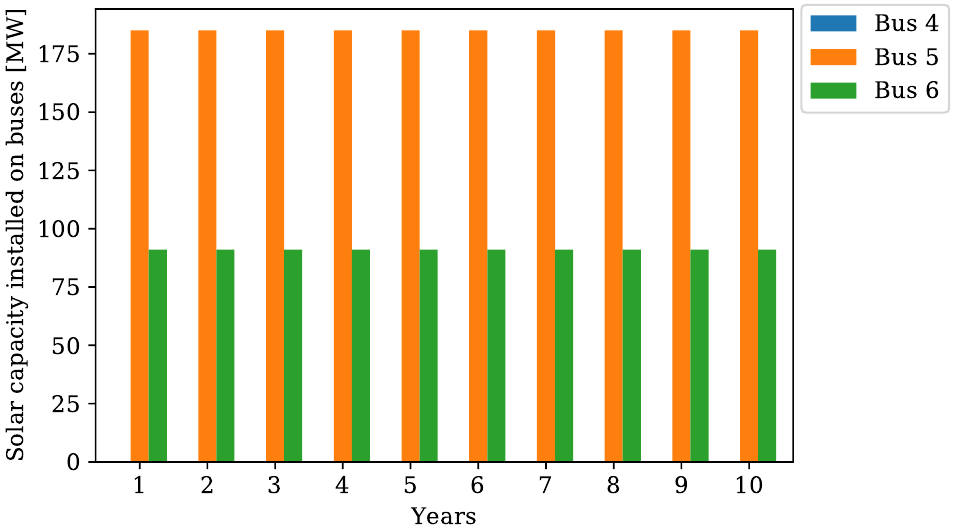}
    \caption{Installed solar capacity on buses during the high wind speed scenario for a 10-year period in the 6-bus test case}
    \label{solar_allocated} 
\end{figure}

The total solar capacity installed during the low and high wind speed scenarios is shown in Table \ref{total_solar_10_year}. During the low wind speed scenario, the solar capacity of 57 MW is installed on bus 5 only throughout the 10-year period.
During the high wind speed scenario, solar energy resources are designated and the distribution on buses is shown in Fig. \ref{solar_allocated}. 
During the 1$^{\text{st}}$ year, 169 MW solar generation installation occurs on bus 5 and it remains the same throughout the 10 years.
78 MW solar generation allocation occurs on bus 6 at the 1$^{\text{st}}$ year, which increases to 88 MW at the 2$^{\text{nd}}$ year due to the increase in demand and lines de-energization.
At the 3$^{\text{rd}}$ year, the installed solar capacity on bus 6 further increases to 91 MW due to increase in conductor clashing score, and it does not change afterwards.

\subsubsection{Load Shedding}
\begin{figure}[bht] 
    \centering
       \includegraphics[width=0.48\textwidth]{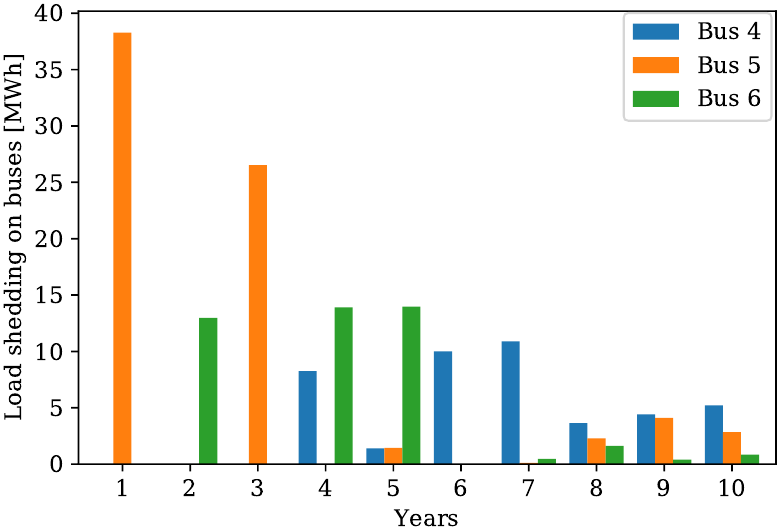}
    \caption{Load shedding on buses during the high wind speed scenario for a 10-year period in the 6-bus test case}
    \label{lost} 
\end{figure}

In this part, the load shedding during the low and high wind speed scenarios is assessed. During the low wind speed scenario, no load shedding occurs because the energized lines are capable of serving the load demand fully. However, load shedding occurs in the high wind speed scenario as shown in Fig. \ref{lost}. It is noticed that no load shedding occurs during 1$^{\text{st}}$ and 2$^{\text{nd}}$ years, because the installed solar generation serves the load demand. During 3$^{\text{rd}}$ year, load shedding of 0.078, 0.86, and 13.39 MWh occur on buses 4, 5, and 6, respectively. Higher load shedding occurs on bus 6 because line $L_7$  is de-energized due to conductor clashing score, and the installed generation capacity is not enough to meet the load demand. At the 6$^{\text{th}}$ year, load shedding decreases due to less fire ignition score and it occurs only on buses 4 and 5 of 2.22 and 5.19 MWh, respectively. From years 7 to 10, the load shedding increases due to higher fire ignition risk. The rise in wind speed leads to greater conductor clashing score, and thus more lines are de-energized which aggravates the amount of load shedding. 

\subsubsection{Modification of Power lines}
Modification of power lines decision variable includes changing the topology of poles or the position of lines to decrease the line's conductor clashing score. During both the low and high wind speed scenarios, no lines are modified because cost of line modification is comparatively higher than the integration cost of of distributed solar generation.


\subsection{IEEE 118-Bus Power System}
In this part, the IEEE 118-bus network is considered for the 10-year expansion planning problem. Due to the large number of power lines in this case, two scenarios are evaluated. In the first scenario, it is considered that 5\% of the lines are subject to conductor clashing, while in the second scenario, 10\% of the lines are at the risk of conductor clashing. The total fire ignition score, net demand, and average wind speed for 10-year period is given in Table \ref{overall_fire_ignition_118_bus_system}.

\begin{table}[bht]
	\vspace{-0.31cm}
	\small \centering
	\caption{\color{black}Total Fire Ignition Score, Net Demand, and Average Wind Speed for 10-year Period in the IEEE 118-Bus System} 
	\vspace{-0.2cm}
	\begin{tabular}{M{0.3cm} M{1.1cm} M{1.05cm} M{1.02cm} M{1.5cm} M{1.5cm}}\hline\hline 
	Years & $\psi_{1-6,y}^{1-9}$ for 5\% of lines & $\psi_{1-6,y}^{1-20}$ for 10\% of lines & Net demand (MW) & Average wind speed for 5\% of lines (mph)& Average wind speed for 10\% of lines (mph)\\ \hline
     1 & 7.26 & 16.31 & 50032 & 61 & 60\\
	 2 & 6.78 & 15.03 & 51032 & 58 & 59\\
	 3 & 7.39 & 15.79 & 52053 & 64 & 60\\
	 4 & 7.66 & 17.27 & 53094 & 65 & 64\\
	 5 & 6.80 & 16.24 & 54156 & 59 & 61\\
	 6 & 8.53 & 18.06 & 55239 & 67 & 66\\
	 7 & 7.23 & 16.63 & 56344 & 60 & 63\\
	 8 & 6.93 & 16.12 & 57471 & 59 & 60\\
	 9 & 7.03 & 16.68 & 58620 & 59 & 62\\
	 10 & 7.07 & 16.02 & 59792 & 60 & 59\\\hline\hline
	\end{tabular}
	\vspace{-0.21cm}
	\label{overall_fire_ignition_118_bus_system}
\end{table}

\subsubsection{Scenario 1: 5\% of the Lines at Risk of Conductor Clashing}

In this part, 5\% of the lines are considered at the risk of clashing. The convergence of CCGA occurs when budget of uncertainty is zero. Here, the uncertainty levels of 10\% and 20\% in the system demand and solar generation are taken into account. With 10\% uncertainty level, CCGA converges in 3 iterations, with a solution time of 9 minutes. In this case, solar generation installations occur on 1-10 buses. The solar generation resource capacity is allocated on buses 1-10 3 for the 10 years. and increases to 299 MW for the remaining 8 years. The increase in solar energy integration occurs due to the increase in load demand. The total operation cost is \$7.691M.
\par
\begin{figure}[bht] 
    \centering
    \includegraphics[width=0.48\textwidth]{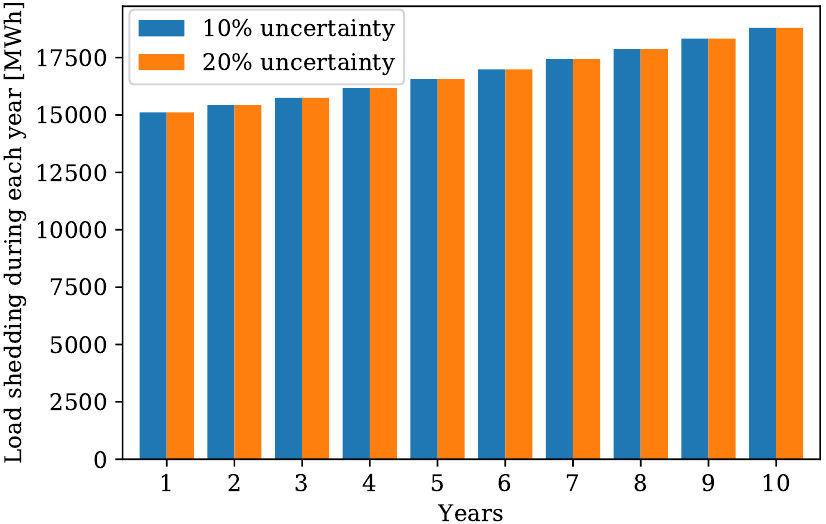}
    \caption{Load shedding during a 10-year period for IEEE 118-bus system when 5\% of the lines are at risk of clashing scenario}
    \label{118_lost_load} 
\end{figure}

When the uncertainty in demand and solar energy generation is 20\%, the algorithm converges in 3 iterations, with solution time of 39 minutes, and the total operation cost of \$8.532M. The less convergence time is because with higher uncertainty level, it is easier to determine the decision variables. The solar energy installation capacity of 297 MW occurs on bus 3 only for the first year and increases to 299 MW for the remaining 9 years.
A comparison of load shedding with 10\% and 20\% uncertainty levels is shown in Fig. \ref{118_lost_load}. The increase in load shedding when moving forward in the planning period is caused by the increase in load demand. The solar energy installation is not able to fulfill the demand and 32 MWh of load shedding occurs on average each year.

\subsubsection{Scenario 2: 10\% of the Lines at Risk of Conductor Clashing}
\begin{figure}[bht] 
    \centering
       \includegraphics[width=0.48\textwidth]{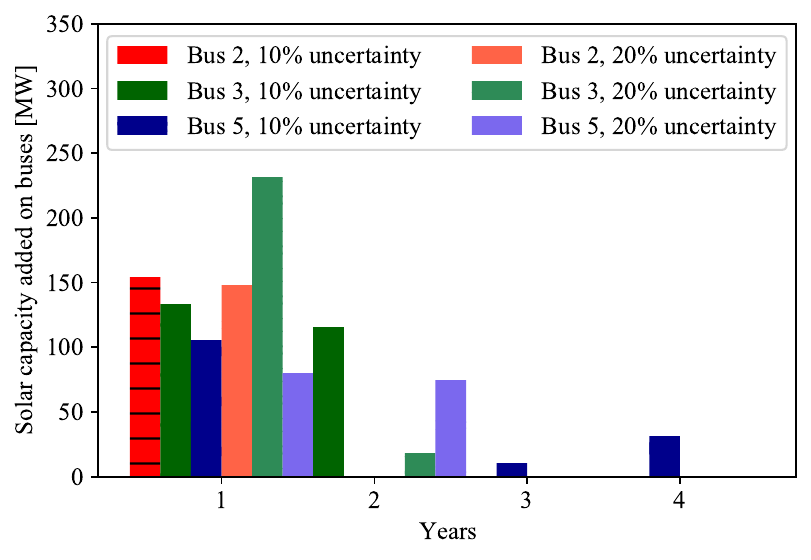}
    \caption{Solar installation on buses with 10\% and 20\% uncertainty levels in solar generation and demand when 10\% of the lines are at risk of conductor clashing in the IEEE 118-bus system}
    \label{solar_118_uncertain} 
\end{figure}

In this scenario, 10\% of the lines are considered at risk of conductor clashing. 
When the uncertainty level in the system demand and the solar generation is 10\%, CCGA converges in 3 iterations, with a solution time of 275 minutes, and the total operation cost of \$8.427M. With 20\% uncertainty level, CCGA converges in 5 iterations, with a solution time of 348 minutes, and the total operation cost of \$9.119M. 
A comparison of solar energy integration on buses 2, 3, and 5 with 10\% and 20\% uncertainty levels is shown in Fig. \ref{solar_118_uncertain}. This figure displays installed capacity for only the first 4 years of the planning horizon, as the installed capacity for the rest of the planning period remains the same as that of year 4. With the increase in uncertainty level from 10\% to 20\%, the operation costs rise as well. In the case of uncertainty level at 10\%, the solar generation integration at bus 3 in the 1$^{\text{st}}$ year is equal to 136 MW, while for the uncertainty level of 20\%, it is increased to 242 MW (78\% increase). During the 2$^{\text{nd}}$ year, the solar installation on buses 2, 3, and 5 is decreased due to more lines de-energization and load shedding.

\par
\begin{figure}[bht] 
    \centering
       \includegraphics[width=0.5\textwidth]{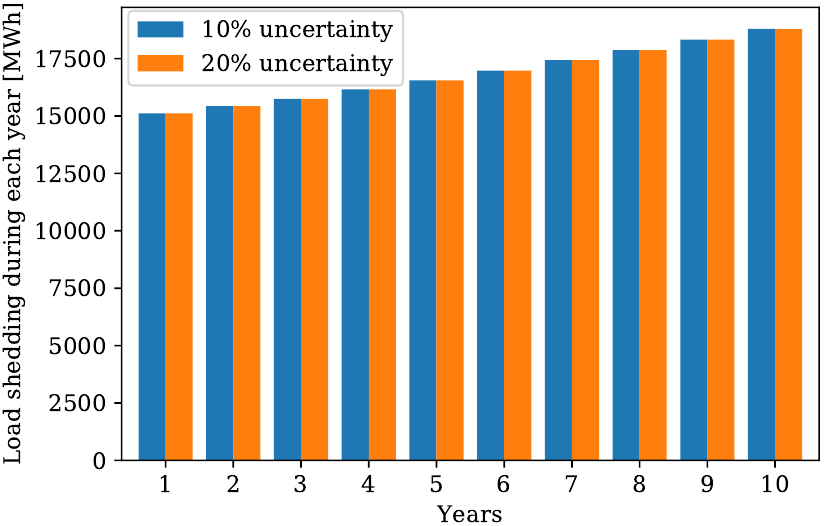}
    \caption{Load shedding during a 10-year period when 10\% of the lines are at risk of conductor clashing in the IEEE 118-bus system}
    \label{118_20_lines_lost_load} 
\end{figure}

The load shedding during a 10-year period is taken into account with 10\% and 20\% uncertainty levels as shown in Fig. \ref{118_20_lines_lost_load}. During the 1$^{\text{st}}$ year, the load shedding amount is equal to 21 MW and it continues to grow for the following years in the planning period as a result of increased yearly demand. At the 10$^{\text{th}}$ year of the operation, the load shedding amount increases to 53 MW in case of 10\% uncertainty level, and to 52 MW in the case of 20\% uncertainty level. 

\subsubsection{Load shedding without Expansion Planning Decision Variables}
\begin{figure}[bht] 
    \centering
       \includegraphics[width=0.5\textwidth]{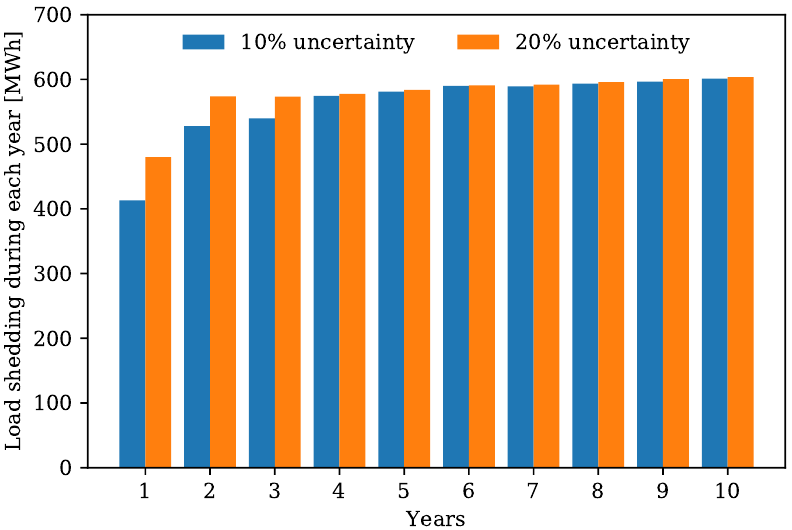}
    \caption{Load shedding during a 10-year period without considering expansion planning decision variables in the IEEE 118-bus system}
    \label{total_lost_load} 
\end{figure}

In this case, it is supposed that no expansion planning decision variables are included. These decisions include distributed solar generation installed capacity, decision for modification of the lines, and the addition of new lines. The load shedding value at a scenario where 10\% of the lines are at risk of conductor clashing is shown in Fig. \ref{total_lost_load}.
 During the 1$^{\text{st}}$ year, the load shedding value in the case of 10\% uncertainty level is 1,850\% higher than that of the case where expansion planning decision variables are taken into account. In the case with 20\% uncertainty level in demand and solar generation, this number is equal to 2,173\%. This illustrates that the expansion planning decision variables are a capable means of reducing the load shedding in a power system at risk of severe weather conditions. 

\section{Conclusions}\label{conclusion}
This paper presented a 10-year expansion planning of electricity grid during severe wildfire weather conditions to find an improved balance for preemptive de-energization of power lines, distributed solar generation, modification of power lines, and addition of new lines. 
A two-stage robust optimization problem is formulated and solved using Column-and-Constraint Generation Algorithm. The first stage minimized the operation cost by considering solar energy integration, addition of new lines, and modification of existing lines while the second stage realized the uncertainty levels in system demand and solar energy generation. The algorithm de-energized lines of high wildfire risk regions and served the customers by integrating distributed solar generation. The impact of severe weather conditions on the energization of power lines and expansion planning cost is assessed. The affect of uncertainty in system demand and solar energy generation is taken into account to depict the real power system scenario. The inclusion of planning decision variables lead to the decrease in load shedding and thus more customers are served.

\ifCLASSOPTIONcaptionsoff
\fi
\bibliographystyle{IEEEtran}

\bibliography{ref}

\end{document}